\documentclass{aa}
\usepackage{graphicx,epsfig}
\usepackage{natbib}
\usepackage[usenames]{color}
\usepackage{txfonts}
\usepackage{url}
\usepackage{stfloats}
\usepackage{fixltx2e}

\newcommand{\kms}{km~s$^{-1}$}

\begin{document}

\title{Break up of returning plasma after the 7 June 2011 filament eruption by Rayleigh-Taylor instabilities}

\author{D.~E. Innes\inst{1} \and R.~H. Cameron \inst{1} \and L. Fletcher \inst{2} \and B. Inhester \inst{1} \and S.~K. Solanki \inst{1,3}}
\institute{Max-Planck Institut f\"{u}r Sonnensystemforschung, 37191 Katlenburg-Lindau, Germany \and School of Physics and Astronomy, SUPA, University of Glasgow, Glasgow, G12 8QQ, UK \and School of Space Research, Kyung Hee University, Yongin, Gyeonggi, 446-701, Korea}
\offprints{D.E. Innes \email{innes@mps.mpg.de}}

\date{Received ...; accepted ...}

\abstract
{A prominence eruption on 7 June 2011 produced spectacular curtains of plasma falling through the lower corona. At the solar surface they created an incredible display of extreme ultraviolet brightenings.}
{To identify and analyze some of the local instabilities which produce structure in the falling plasma.}
{The structures were investigated using SDO/AIA 171\AA\ and 193\AA\ images in which the falling plasma appeared dark
against the bright coronal emission.}
{Several instances of the Rayleigh-Taylor instability were investigated. In two cases the Alfv\'en velocity associated with the dense plasma could be estimated from the separation of the Rayleigh-Taylor fingers.
A second type of feature, which
has the appearance of self-similar branching horns was discussed.}
{}


\titlerunning{Break up of returning filament plasma}
\authorrunning{Innes, Cameron, Fletcher, Inhester, Solanki}

\maketitle

\section{Introduction}

One of the most spectacular solar events seen so far with the Atmospheric Imaging Assembly (AIA)
on the Solar Dynamics Observatory (SD0) occurred when a filament erupted on 7 June 2011\footnote{http://www.thesuntoday.org/current-observations/a-spectacular-event-a-filamentprominence-eruption-to-blow-your-socks-off/}. The eruption that was associated with a fast coronal mass ejection (CME), an M2 flare and dome-shaped extreme ultraviolet (EUV) front \citep{cheng12}, slung material across almost a quarter of the solar surface.
The 'S' shaped filament erupted from the southern active region AR11226 as it was nearing the western limb. The eruption started at 6:00~UT, and reached its peak GOES soft X-ray brightness at 6:35~UT.
Non-escaping material was seen in SOHO/LASCO C2
and STEREO/COR1
images falling back from heights up to 4 solar radii.
 The first returning material was seen at the solar surface at around 7:00~UT in SDO/AIA and STEREO/EUVI-A images.

\begin{figure*}
\centering
\includegraphics[width=13cm]{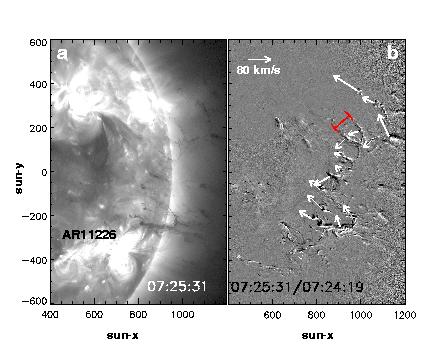}
\caption{Falling (dark) plasma after the 2011 June 7 filament eruption looks like the fingers on the rim of a splash: a) 193\AA\ intensity; b) 193\AA\ ratio of images at the times given (see Sect. 2 for details). The white arrows in (b) represent plane-of-sky velocities and the red bar the typical spacing between fingers.} \label{crown}
\end{figure*}

Here
 we concentrate on the structure of the falling plasma. An hour and a half after the eruption it looked like a huge upside-down crown extending over at least 600\arcsec\ (Fig.~\ref{crown}).
The leading edge had broken-up into semi-regular arcs and spikes, similar to the rim of a splash \citep[e.g.][]{Edgerton87}.
Such long fingers and arcs are also seen in the Crab supernova remnant \citep{Hester08}.
 In the Crab, the main fingers are thought to be the result of the magnetic Rayleigh-Taylor (RT) instability \citep{Hester96}. RT has also been invoked to explain filamentary structure
\citep{Isobe05} and prominence bubbles \citep{Berger11} on the Sun.

 In this letter we highlight features that lead us to conclude that the magnetic RT is also responsible for the structures seen in this event.
 Further we investigate properties of the magnetic RT instability to obtain diagnostics of the local plasma conditions.

\section{Observations}
SDO/AIA takes images
 of the full solar disk with a resolution of about 0.6\arcsec\ pixel$^{-1}$ through 10 filters, selected
to single out specific strong lines in the corona and continuum
emission from the lower chromosphere. We investigate the structure of cold filament plasma which was seen as dark structures in the 171, 131, 193, 211, 304, and 335\AA\ images due to absorption of background EUV emission by neutral hydrogen, helium and
singly ionized helium. Since we will be discussing the structure of the plasma as it appears against the coronal background, we only show 171\AA\ and 193\AA\ data because these have the best contrast. The data are presented as either intensity or ratio images. The ratio images are the log of the ratio of the image at the time given and the image taken 12~s earlier except where times are given specifically. In the ratio images the absorbing plasma has moved from the bright to the dark regions. Flow velocities have been computed with the optical flow code of \citet{Gissot07} and are represented by white arrows in the ratio images.

\subsection{Onset}
The overall development of the filament eruption is shown in Fig.~\ref{onset_crown}. Fig.~\ref{onset_crown}a gives the impression of a hot core with a cool cone-like shell expanding upwards above it.
There is also a cloud of hot ejecta coming from the center (red arrow). Focusing on the northern edge of the dark cone, we see that initially this edge was essentially straight. Then, as the erupting material expanded, parts broke away (red box in Fig.~\ref{onset_crown}b) and the upper edge started to corrugate. These corrugations later developed into the finger and arc structure in Fig.~\ref{crown}. Unfortunately they were too far off limb to clearly observe their evolution in the SDO/AIA images due to low contrast with the background.

\subsection{Spikes}
\begin{figure}
\centering
\includegraphics[width=8.6cm]{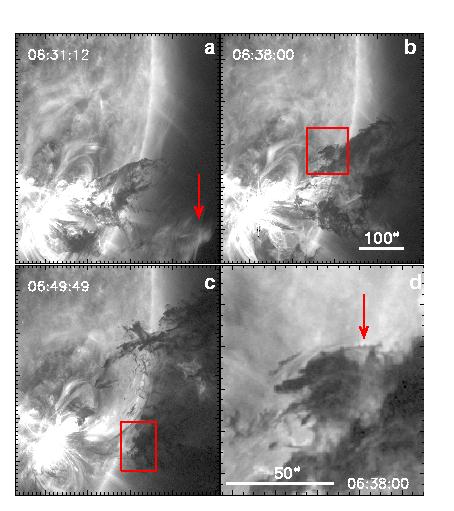}
\caption{Erupting filament plasma seen in 171\AA\ intensity images. The red arrow in (a) points to hot ejecta. Details of the red box in (b) are shown in (d). The red box in (c) outlines the regions displayed in Figs.~\ref{rt_193} and ~\ref{closeup}. In (d) the arrow points to the spikes shown in detail in Fig.~\ref{kh_rt}. The FOV is 500\arcsec x 550\arcsec\ except in (d) where it is 96\arcsec x 110\arcsec.} \label{onset_crown}
\end{figure}

\begin{figure}
\centering
\includegraphics[width=5cm]{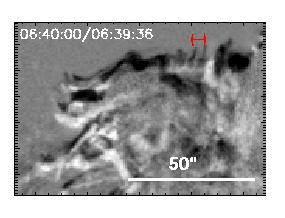}
\caption{Growth of spikes along filament indicated by the red arrow in Fig.~\ref{onset_crown}d: 171\AA\ intensity ratio image. The red bar indicates the spikes' separation. The FOV is 96\arcsec x 66\arcsec.} \label{kh_rt}
\end{figure}

The part that broke away shows small regularly spaced spikes pointing outward along its edge (Fig.~\ref{onset_crown}d, red arrow).
They look similar to the spikes observed by SDO/AIA in the 131\AA\ filter on the edge of a coronal eruption and interpreted by \citet{Foullon11} as Kelvin-Helmholtz roll-ups. The ones here appear to be different because, as implied by the ratio image of this filament (Fig.~\ref{kh_rt}) and shown in the online appendix (Fig.~\ref{appfig}), they grew upwards and did not turn over. It is interesting to note that the growth which was along the flow direction was perpendicular to the solar radial direction, hence not governed by gravity.

The spikes in Fig.~\ref{kh_rt} had a separation of 5\arcsec, length of 6\arcsec, and growth rate 12~\kms.
About 1~min later they faded and were overtaken by plasma from behind. Similar spikes were seen along the edge of filamentary structures throughout the evolution and they were all pointing in the direction of motion and disappeared without turning over.

\subsection{Fingers and Arcs}
\begin{figure}
\centering
\includegraphics[width=\linewidth]{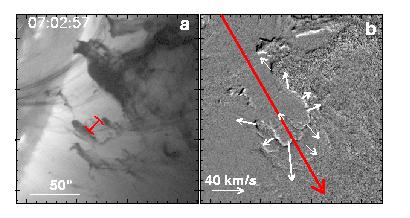}
\caption{Expanding arc with small-scale fingers and arcs inside: (a) 193\AA\ intensity (b) 193\AA\ ratio image. The red bar in (a) spans an arc between two small-scale fingers. The white arrows represent plane-of-sky velocity and the red line shows the position of the time slice in Fig.~\ref{rt_time}. The FOV is 180\arcsec x 185\arcsec. The evolution can be seen in the movie 
{\tt arcs\_movie}} \label{rt_193}
\end{figure}

The falling filament plasma closest to the active region presented a sequence of well-exposed images showing
arcs and some small-scale fingers. A series of processes occurred. Structure in the ejected plasma was initially stretched out
by its large-scale expansion, creating small cavities that then expanded by pressure gradients probably associated with outflow from the eruption site. As the cavity expanded, new smaller-scale fingers and arcs formed on the inside. The movie,
{\tt arcs\_movie} shows the complete evolution of the cavities and
small-scale fingers and arcs on the inside. A snapshot of the movie is shown in Fig.~\ref{rt_193}a. The red bar bridges one of these small-scale arcs. The outline of an expanding cavity is clearly visible in the ratio image (Fig.~\ref{rt_193}b).

\begin{figure}
\centering
\includegraphics[width=8.6cm]{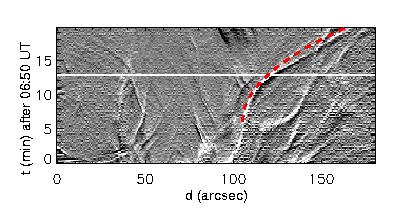}
\caption{Upwards expansion of arc: space-time slice of 193\AA\ running ratio images along the red line in Fig.~\ref{rt_193}b. The best fit $\mathrm{d} \propto \mathrm{t}^2$ (Eq.~3) relationship is shown as a red dashed line. A white line is drawn at the time of the image in Fig.~\ref{rt_193}} \label{rt_time}
\end{figure}

To study the acceleration of the arcs, we show the space-time running ratio image in Fig.~\ref{rt_time}. It was
taken along the red arrow drawn through the apex of the arc in Fig.~\ref{rt_193}b.
The expanding edge of the arc is over-plotted with the red line $d\propto t^2$ which is the relationship expected for constant
acceleration of the interface. Space-time images along different directions through the arc give similar results but the edge is
not as sharp. This is the background on which the small-scale RT fingers we are interested in develop.

\subsection{Horns}

\begin{figure}
\centering
\includegraphics[width=8.6cm]{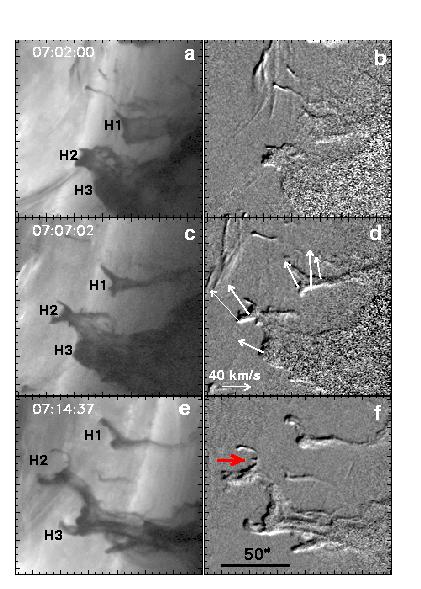}
\caption{Formation of horns: (left) 171\AA\ intensity; (right) 171\AA\ ratio. White arrows represent plane-of-sky velocity. FOV is 130\arcsec x120\arcsec. The structures are labelled H1, H2, H3. The red arrow in (f) points to a secondary spike. The evolution of these structures is shown in the movie, 
{\tt horns\_movie}.} \label{closeup}
\end{figure}

Another set of features we would like to point out are the horns. The formation of three of these is illustrated
in Fig.~\ref{closeup} where they are labelled H1, H2, and H3, and in the accompanying movie, 
{\tt horns\_movie}. These horns sometimes formed out of sheets that distorted
(H1 and H2) and sometimes when a thread tore away from the main stem (H3).
 At the time the central frame was taken, H1 was compressing as it moved northward as though being pushed from the south.
Most the other structures are also moving north although the tips of the the denser/darker fingers (e.g. H2, H3) have a
significant sunward component. These ones later developed arc-like horns. In Fig.~\ref{closeup}f, a small
spike has developed inside H2 (red arrow), repeating in a self-similar way the finger and arc structure.



The movie
shows
there were several background effects that could have been influencing the evolution and growth of the horns because (i)
nearly all the structures are moving northward, (ii) there are hot outflows from the eruption site, and (iii) at the end of
the movie there is a bright ridge off-limb along which dark plasma seems to be falling. The ridge and sideways motion can be
attributed to the coronal magnetic field configuration which at the height of the falling plasma consisted of closed loops
connecting the active regions north and south of the equator (Fig.~\ref{crown}).
There are no obvious signs of a direct influence of the outflows on the evolution of the falling plasma, but the flows would
probably have been directed along field lines connecting the active regions and could therefore have added momentum to
the filament motion. The horns might thus be associated with the global field configuration rather than being a purely local
phenomenon.

\section{Discussion}
Since both the time and spatial scales are large, the evolution of the instabilities studied here
can be treated as MHD phenomena. The examples discussed in Sect. 1, 2.1, 2.3 are all cases of cool
dense material sitting on top of lighter hotter plasma.
In the features discussed in 2.2 and 2.3, a large-scale pressure
gradient appears to be accelerating the structure (see Fig.~\ref{rt_time} where constant acceleration of
the interface can be readily inferred). The RT instability occurs whenever a denser fluid is accelerated against a less dense
fluid by, for example, gravity or pressure gradients. So if we ignore the effects of the magnetic field,
each of these cases is RT unstable.
Indeed the large-scale structure of the falling filament plasma looks like the rim of a large splash, in
which long fingers are connected by shadowy arcs (Fig.~\ref{crown}). Although the micro-physics is different it is
probable that here, like in the splash \citep{Allen75}, RT instabilities were responsible for the
break-up.

To proceed further we need to consider the structure of the magnetic field in each particular case. Although the dense, cool filament plasma is partially neutral there will be strong collisional coupling between the neutrals and the ionized component. It is also reasonable to assume that
 the hot and cold plasma lie on different field lines because thermal conduction is efficient along field lines. In this case the magnetic
field must be parallel to the interface between the two plasmas.
Magnetic tension then tends to inhibit transverse motions which vary along any field line, with shorter wavelength fluctuations
being more strongly inhibited. For the case where the magnetic field in both fluids is oriented in the same direction,
the instability only occurs for waves which have $k_{\parallel}<k_c$ where $k_{\parallel}$ is the component of
the wave vector aligned with the magnetic field \citep{Chandrasekhar61}. The wavelength associated with $k_c$ is
\begin{equation}
\lambda_c = {{B^2}\over{g(\rho_h - \rho_l)}}
\label{eqn:lambda}
\end{equation}
where $\rho_h/\rho_l > 1$ is the ratio of the densities on the two sides of the interface,
$B$ is the strength of the uniform magnetic field, and $g$ is the net acceleration
due to gravity and additional acceleration, $a$, caused by pressure
gradients across the interface.

Modes with $k_{\parallel}=0$ are unaffected by the field, with the consequence that the instability acts to form sheets aligned
with the field \citep{Isobe05}.
Here, however, the magnetic field orientation in the cold plasma and in the
warm plasma will not, in general, be aligned. In this case the magnetic tension acts against the instability
for all ${\bf{k}}$.

Simulations of the magnetic RT instability with different magnetic field orientations on the two sides of
the density jump have been performed by \citet{Stone07}.
If the fields on either side of the interface are oriented parallel to the interface but at an angle to each other (the case we
are concerned with), bubbles separated by long fingers on the scale of the critical wavelength grow preferentially. Their simulation produced long fingers separated by smooth arcs, morphologically similar to those in Fig.~\ref{crown}.
The spacing of the RT bubbles corresponds to $\lambda_c$ as described above. Hence by measuring the separation of the
fingers we can use Eq.~\ref{eqn:lambda} to place constraints on the plasma properties.

For all our examples the heavy plasma is much denser than the light plasma, so we take $\rho_h \gg \rho_l$,
and using $B^2 = 4\pi\rho_hV_A^2$, where $V_A$ is the Alfv\'en speed of the falling filament material where the instability starts, we obtain
\begin{equation}
V_A= \sqrt{\lambda_c g/(4\pi)}.
\label{eqn:inv}
\end{equation}
\noindent
Thus the observed 100~Mm separation of the fingers seen in Fig.~\ref{crown} gives an Alfv\'en speed of
approximately 47~\kms\ for the falling material. We note that since we can measure the spacing between
the fingers only in the plane perpendicular to the line-of-sight, this is a lower limit to the Alfv\'en velocity.

A slightly more complicated example is that corresponding to Figs.~\ref{rt_193} and \ref{rt_time}.
Here acceleration due to a large-scale pressure gradient is important. The acceleration of the interface can be
measured by fitting
\begin{equation}
 d = \frac{a}{2} t^2
\end{equation}
\noindent
to the location of the interface shown in Fig.~\ref{rt_time}, where $d$ is the distance moved by the interface, $t$ is time, and $a$ is the
constant acceleration, which we measure to be 140~m~s$^{-2}$. It is directed away from the Sun, so increases the effective $g$ which appears in
Eq.~\ref{eqn:inv}. Based on the observed spacing between the main fingers, 10~Mm, in the plane perpendicular to the
line-of-sight, the lower limit for the Alfv\'en velocity in the dense plasma is 18~\kms.
Throughout many smaller fingers appear but these are quickly damped, presumably because the spacing is less than the critical wavelength.

The spikes of Sect~2.2 seem to be also produced by the RT instability based on their apparent evolution. We suspect that they are a consequence of the blast, seen 10~min earlier as a dome-like EUV wave \citep{cheng12}, that overtook the filament plasma.
The observed short wavelengths are consistent with rapid acceleration. After the acceleration the RT instability will cease and magnetic tension will act to remove the fingers.

Lastly we note that there are certainly other instabilities besides the RT which could be analyzed. The above examples were
chosen because they were relatively clean, simple examples where the entire evolution could be followed. The horns discussed in
section 2.4 are presumably the result of some instability which we have not been able to properly identify.
We see no examples of Kelvin-Helmholtz even along the edges of the fingers where they could be expected in the non-linear phase of the RT instability.

\section{Conclusion}
In this letter we have analyzed spatially and temporally localized instabilities associated with
the event on June 7 2011. Concentrating on examples of the RT instability, we have shown that reasonable values for the Alfv\'en velocity in the falling plasma can be derived.

\begin{acknowledgements}
The authors are indebted to the SDO/AIA teams and the German Data Center at MPS for providing the data. This work has been supported by WCU grant No. R31-10016 funded by the Korean Ministry of Education, Science and Technology and by grant ST/1001808/1 from UKs Science and Technology Facilities Council, and Leverhulme Foundation Grant F00-179A.
\end{acknowledgements}

\bibliographystyle{aa}


\begin{appendix}

\section{Spikes' evolution}
\begin{figure*}[ht]
\centering
\includegraphics[width=14cm,clip]{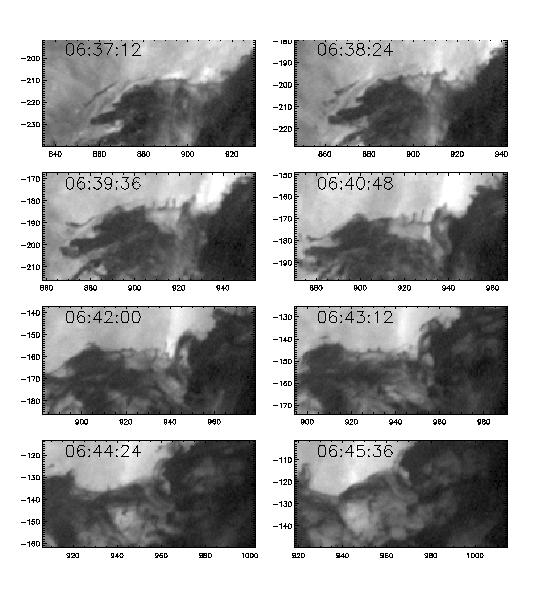}
\caption{Series of 171\AA\ snapshots showing the evolution of the spikes described in section 2.2 and Fig~\ref{kh_rt}. Small cavities develop between the spikes (see images at 06:24:48 and 06:26:00) while they continue to grow without rolling over. After only a few minutes the spikes fade and are overtaken denser plasma at their base.}
\label{appfig}
\end{figure*}
\end{appendix}

\end{document}